\documentclass[aps,pra,preprint,groupedaddress,showpacs]{revtex4}

\usepackage{graphicx}
\usepackage{verbatim}
\usepackage{amsfonts} 
\usepackage{amsmath}
\usepackage{mathbbol}
\usepackage{amssymb}   
\usepackage{color}
\usepackage{bm}

\newcommand{\be}{\begin{equation}}
\newcommand{\ee}{\end{equation}}
 
 \definecolor{BrickRed}{cmyk}{0,0.89,0.94,0.28}
\definecolor{MidnightBlue}{cmyk}{0.98,0.13,0,0.43}
\definecolor{DarkGreen}{rgb}{0,0.7,0.1}

\begin{document}

\title{Surface Scattering Expansion for the Casimir-Polder Interaction of a Dielectric Wedge}

\author{Thorsten Emig}

\affiliation{Laboratoire de Physique
Th\'eorique et Mod\`eles Statistiques, CNRS UMR 8626,
Universit\'e Paris-Saclay, 91405 Orsay cedex, France}

\email{thorsten.emig@cnrs.fr}

\date{\today}

\begin{abstract}
The electromagnetic scattering amplitude of a dielectric wedge is not known in closed form. This makes the computation of the Casimir-Polder (CP) interaction between a polarizable particle and a dielectric wedge challenging. This geometry is a prototype for the effect of edges on fluctuation-induced interactions, and hence it is important to employ new methods for this problem. Using a recently developed multiple scattering expansion [T.~Emig and G.~Bimonte, Phys.~Rev.~Lett.~{\bf 130}, 200401 (2023)], here we implement a basis-free numerical evaluation of this expansion to obtain precise estimates of the CP potential for a wedge over a wide range of dielectric constants. A remarkable finding is that the CP potential for a dielectric wedge with a smoothed edge is closely related to the potential of a sharp wedge made of a perfect electric conductor. The latter potential is known exactly, making this relation particularly useful in practice.
\end{abstract}


\maketitle

\section{Introduction}

In many fields of physics, biology and chemistry one commonly observes small particles, molecules or atoms in close vicinity to the surface of macroscopic bodies. Even in the absence of charges, the particles experience a 
long-ranged surface interaction which can be viewed as resulting from modifications of the quantum and thermal fluctuations of the electromagnetic field by the bodies. Casimir and Polder \cite{Casimir:1948jf} examined this interaction between a neutral atom and a flat conducting plate, demonstrating that the altered vacuum generates a spatially varying Lamb shift. The gradient of this shift results in an attractive long-range force, referred to as Casimir-Polder (CP) force.
After Casimir's earlier pioneering discovery that two uncharged, perfectly conducting parallel plates at zero temperature experience an attractive force due to quantum fluctuations of the electromagnetic field \cite{casimir}, Lifshitz extended this concept by applying the then-novel method of fluctuational electrodynamics. He calculated the Casimir force between two parallel, infinite surfaces of dispersive and dissipative dielectric materials at finite temperatures \cite{lifshitz}. By considering the dilute limit of one body, Lifshitz also derived the CP force between a small polarizable particle and a planar surface. His result remained the only one for a long time, as extending the computations beyond simple planar geometries proved challenging. The complexity arises from the collective and non-additive nature of dispersion forces. For many years, the only way to approximate dispersion forces in non-planar configurations was through the proximity approximation \cite{derjaguin}, for in-depth discussions see \cite{parsegian,galina,buhmann1,rodriguez,woods,bimonte2017}. 

The first clear experimental evidence for the existence of the CP force was obtained for sodium atoms transmitted through a wedge-like cavity \cite{Sukenik:1993aa}. The cavity consisted of gold surfaces, and the measured intensity of the transmitted atom beam was found to be in good agreement with the theoretical prediction of the CP force for perfect electric conductors (PEC). However, at  smaller cavity widths a trend towards a reduced force due to imperfect reflectivity of the gold cavity was observed. 
More recently, numerous CP force measurements have been performed using various techniques. These methods include ultra-cold atom experiments \cite{marachevsky2014}, spectroscopy with nano-sized cells \cite{Peyrot:2019aa,Whittaker:2014aa,Carvalho:2018aa}, and atomic diffraction through material nano-gratings \cite{Lonij:2009aa,Garcion:2021aa,Morley:2021aa}. The force measurements involve also non-metallic materials with complex shapes, and hence a better theoretical understanding of the CP force beyond planar surfaces and PEC is required. 

Significant theoretical advancements in understanding Casimir forces between macroscopic bodies were achieved with the development of the scattering approach \cite{emig2007, kenneth2008, rahi2009}, originally formulated for non-planar mirrors \cite{Genet03, lambrecht}. This method describes the interaction between dielectric bodies using their scattering amplitude, also known as the T-operator. Although this approach has driven much of the recent theoretical progress, the T-operator is only explicitly known for highly symmetric objects, like spheres and cylinders, or for certain perfectly conducting geometries \cite{maghrebi}.
A more fundamental limitation of the scattering approach arises for interlocked geometries, as the method fails due to the lack of convergence of the mode expansion \cite{Wang2021}. The need for theoretical methods capable of accurately computing forces in complex geometries has become increasingly pressing, particularly because recent experiments with micro-fabricated surfaces \cite{Banishev:2013zp, Intravaia:2013yf, Wang2021} have revealed significant deviations from the proximity  approximation. Some theoretical advancements have been made for the specific case of dielectric rectangular gratings, utilizing a generalized Rayleigh expansion \cite{marachevsky2008, marachevsky2014, marachevsky2020}. Additionally, a general approach has been developed for gently curved surfaces, where a gradient expansion can be applied to derive first-order curvature corrections to the proximity approximation for the Casimir force \cite{fosco, bimonte2012, bimonte2012bis, bimonte2017bis}. More recently, it has been demonstrated that surface integral-equation methods \cite{chew, volakis}, long used in computational electromagnetism, can also be applied to calculate Casimir interactions for arbitrary configurations of homogeneous magneto-dielectric bodies of any shape. The formulation in 
\cite{reid2013} represents Casimir potentials as trace of a surface operator. However, a potential challenge with this approach is that it involves the inverse of the surface operator. Hence a boundary  element method (BEM) is needed to replace the surface operator with a matrix, where each element is a double surface integral of the free Green’s tensors over pairs of the surface elements. This method is computationally intensive, also due to the strong inverse-distance cubed singularity of the surface operator in the coincidence limit.

Again based on surface integral equations, recently  a multiple scattering expansion (MSE)  of Casimir and CP interactions for dielectric bodies of arbitrary shape has been derived \cite{emig2023,bimonte2023,Bimonte:2024aa}.  
An important property of this approach is that the Casimir potential is expressed as the trace of a Fredholm integral operator of the {\it second} kind. This allows for the expansion of the Casimir potential in powers of surface operators. These operators are explicitly known in terms of the free Green tensor of the electromagnetic field in a homogeneous medium.
The expansion can be physically understood as a series in the number of scatterings occurring on the surfaces of the bodies. Compared to the scattering approach, the MSE has the advantage that no knowledge of the scattering amplitude (the T-matrix) of the bodies is required, and hence no partial wave basis needs to be introduced. In comparison to the numerical BEM approach, the MSE has the advantage that no surface basis functions (for the surface currents) are required since the surface operator (in the least divergent formulation, see below) can be integrated in the ordinary sense over the surfaces of the bodies. 

The CP potential between a particle and a macroscopic body is usually easier to compute than the Casimir interaction between macroscopic bodies as the particle can be considered as point-like, hence probing the vacuum fluctuations only at a localized position in space. This problem has been studied by many authors in the past, using a variety of methods.
For spheres and cylinders, the scattering approach leads to  exact expression for the CP potential in terms of the T-operator \cite{Marachevsky2002,Buhmann2004,Milton2012,Noruzifar:2012wk}. A Rayleigh expansion  has been used for the CP interaction between an atom and a rectangular dielectric grating  \cite{Contreras2010,Bender2014}. It has been shown  that  a gradient expansion    can be  used  to compute   the leading and the next-to-leading curvature corrections beyond the proximity approximation for the CP energy for a gently curved surface \cite{bimonte2014,bimonte2015}.  A numerical time-domain approach to compute the CP interaction of an atom with an arbitrary micro-structured body  has been  discussed recently \cite{Intravaia2023}.

In this work, we investigate the CP potential for a surface with an gently smoothed edge within the MSE approach.
We study a wide range of dielectric contrasts, and compare our findings to the exactly known CP potential for sharp a PEC wedge, and to the proximity approximation. For the first time, we implement here a completely surface basis-free computation of the MSE, involving only ordinary numerical surface integration of tangential components of the free Green tensor for homogeneous media.  We demonstrate that 2nd order MSE combined with a non-linear convergence acceleration method leads to precise estimates of the CP potential.

\section{Multiple Scattering Expansion for the Casimir-Polder potential} 
\label{sec:MSE}

\begin{figure}[h]
\includegraphics[width=0.55\textwidth]{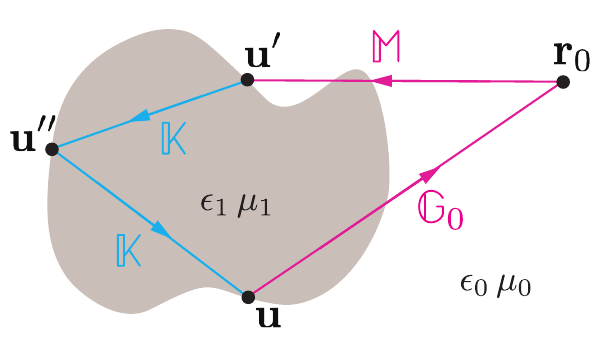}
\caption{Configuration of a dielectric body with electric and magnetic permeabilities $\epsilon_1$, $\mu_1$, interacting with a polarizable particle outside the object. The electric and magnetic permeabilities of the surrounding medium are $\epsilon_0$, $\mu_0$. The operators of the MSE are defined on the body's surface (surface operator $\mathbb{K}$), acting between surface points ${\bf u}$, ${\bf u}'$ and ${\bf u}''$. The operators $\mathbb{M}$ and $\mathbb{G}_0$ connect the surface points to the location $\bf{r}_0$ of the particle. The order of the MSE is determined by the maximum number of operators $\mathbb{K}$ included on a path (here shown for order two). See text for details. 
\label{fig:1}}
\end{figure}

Here we briefly review the MSE for the Casimir-Polder (CP) potential of a small polarizable particle in the presence of a dielectric body with permittivities $(\epsilon_1$, $\mu_1)$, surrounded by a medium with  permittivities $(\epsilon_0$, $\mu_0)$. For a detailed description and derivation see Refs.~\cite{emig2023,bimonte2023}. To focus on the essential geometric aspects of the interaction, we assume that the particle is only electrical polarizable with a frequency independent and isotropic (scalar) polarizability $\alpha$. At zero temperature, the CP potential is then determined by only the electric component (EE) of scattering Green tensor $\Gamma({\bf r},{\bf r}')={\mathbb G}({\bf r},{\bf r}')-{\mathbb G}_0({\bf r},{\bf r}')$ where ${\mathbb G}$ is the electromagnetic (EM) Green tensor in the presence the body and ${\mathbb G}_0$ the free EM Green tensor for a homogeneous space filled by a medium with permittivities $(\epsilon_0$, $\mu_0)$. The latter tensor depends on frequency, which can be integrated along the imaginary frequency axis to yield the potential
\begin{equation}
U({\bf r}_0) = -2\,  \hbar c \, \alpha \int_0^\infty d \kappa \, \kappa \, \Gamma^{(EE)}({\bf r}_0,{\bf r}_0) 
\end{equation}
with the reduced Plack's constant $\hbar$ and the speed of light $c$. In general, a closed form expression for $\Gamma$ is not known. Therefore, approximations have to be employed.  Using the surface integral-equation formulation for EM scattering by a dielectric body, the scattering Green tensor can be expressed as a double surface integral over the surface $S$ of the dielectric body,
\begin{equation}
\label{eq:Gamma}
{\Gamma}({\bf r},{\bf r}')=
\int_S ds_{\bf u} \int_S ds_{{\bf u}'}\,  \mathbb{G}_0({\bf r},{\bf u}) (\mathbb{I}-\mathbb{K})^{-1}({\bf u},{\bf u}') \mathbb{M}({\bf u}',{\bf r}')\;.
\end{equation}
Here $\mathbb{K}$ is the central object of the surface formulation. It is a surface scattering operator, defined for two surface points ${\bf u}$, ${\bf u}'$ by
\begin{equation}
\label{eq:operator_K_general}
\mathbb{K}({\bf u},{\bf u}') = 2 \mathbb{P} (\mathbb{C}^{i}+\mathbb{C}^{e})^{-1} {\bf n}({\bf u}) \times  \left[ \mathbb{C}^{i}  \mathbb{G}_1({\bf u},{\bf u}')- \mathbb{C}^{e} \mathbb{G}_0({\bf u},{\bf u}')\right] \, , \quad \mathbb{P} = \big(\begin{smallmatrix} 0 & -1 \\ 1 & 0 \end{smallmatrix} \big)\, ,
\end{equation}
and $\mathbb{M}({\bf u},{\bf r})$ is the bulk-surface operator
\begin{equation}
\label{eq:operator_M}
\mathbb{M}({\bf u},{\bf r}) =  -2 \mathbb{P} (\mathbb{C}^{i}+\mathbb{C}^{e})^{-1} \mathbb{C}^{e}\, {\bf n}({\bf u}) \times  \mathbb{G}_0({\bf u},{\bf r}) 
\end{equation}
which describes the wave propagation from a point in the bulk space outside the body to the first point on the surface, see Fig.~\ref{fig:1}. 
In the above equations, $\mathbb{G}_1$ is the free Green tensor for the homogeneous media with permittivities $(\epsilon_1$, $\mu_1)$, respectively (see App.~E of \cite{bimonte2023} for explicit expressions for $\mathbb{G}_0$ and $\mathbb{G}_1$),
while ${\bf n}({\bf u}) $ is the outward unit normal vector to the surface $S$ at point ${\bf u}$ \footnote{The action of ${\bf n}({\bf u}) \times$ on the $3 \times 3$ matrices $\mathbb{G}^{(pq)}_1$ and $\mathbb{G}^{(pq)}_{0}$ ($p,q \in \{ E,H\}$) are respectively defined by $({\bf n}({\bf u}) \times \mathbb{G}^{(pq)}_1){\bf v} \equiv {\bf n}({\bf u}) \times (\mathbb{G}^{(pq)}_1 {\bf v})$ and  $({\bf n}({\bf u}) \times \mathbb{G}^{(pq)}_{0}){\bf v} \equiv {\bf n}_\sigma({\bf u}) \times (\mathbb{G}^{(pq)}_{0} {\bf v})$,  for any vector ${\bf v}$.}. 
It is important to  note that $\mathbb{K}$ and $\mathbb{M}$ depend on  $4$  {\it arbitrary} coefficients, which must form two invertible diagonal $2\times 2$ matrices $\mathbb{C}^{i}$, $\mathbb{C}^{e}$.  We note that only the projections of $\mathbb{K}({\bf u},{\bf u}')$ on the tangential spaces at ${\bf u}$ and ${\bf u}'$ contribute to the scattering Green function and the Casimir energy as the images of $\mathbb{M}$ and of $\mathbb{K}$ itself lie in the tangential space. 

This surface formulation has two major advantages over previous approaches for Casimir interactions: \\
(1) an unpleasant feature of the  operator $\mathbb{K}$ is that in general it has a singularity $\sim 1/|{\bf u}-{\bf u}'|^\gamma$ with $\gamma=3$ when the two surface positions ${\bf u}$, ${\bf u}'$ approach each other. Interestingly, there exists a choice of coefficients for which the singularity is reduced to a weaker divergence with exponent $\gamma=1$, at least for a {\it smooth} surface. This choice corresponds to
\begin{equation}
\label{eq:coefficients}
\mathbb{C}^{i}={\rm diag}(\epsilon_1,\mu_1)\, , \quad
\mathbb{C}^{e}={\rm diag}(\epsilon_0,\mu_0) \, .
\end{equation}
The corresponding surface operator $ \mathbb{K}$  has unique mathematical properties \cite{muller,bimonte2023}. \\
(2) The Fredholm type of the operator $(\mathbb{I}-\mathbb{K})^{-1}$ in Eq. (\ref{eq:Gamma}) permits an expansion of ${\Gamma}({\bf r},{\bf r}')$ in powers of $\mathbb{K}$. Hence, no direct inversion of this operator is required. Instead, the inverse operator can be obtained as a Neumann series, leading to a MSE for the CP potential in terms of the number of scatterings at the surface of the body. For the first two orders, this expansion reads (for an electrically polarizable particle)
\begin{align}
\label{eq:CP_1st_order}
U({\bf r}_0) &= - 2 \,  \hbar c \,  \alpha \int_0^\infty d \kappa \, \kappa \left\{ \sum_{p={E,H}} \int_S ds_{\bf u}  {\rm tr} \left[\mathbb{G}_{0}^{(E p)}({\bf r}_0,{\bf u})    \mathbb{M}^{(p E)}({\bf u},{\bf r}_0) \right] \right. \nonumber \\
&\left.+\sum_{p,q={E,H}} \int_S ds_{\bf u}  \int_S ds_{{\bf u}'}  {\rm tr} \left[\mathbb{G}_{0}^{(E p)}({\bf r}_0,{\bf u})   \mathbb{K}^{(p q)}({\bf u},{\bf u}')   \mathbb{M}^{(q E)}({\bf u}',{\bf r}_0) \right] \right\}+\cdots 
\end{align}
and higher orders are given simply by surface integrals over higher powers of $\mathbb{K}$.
The weaker singularity of $\mathbb{K}$ with the choice of Eq.~(\ref{eq:coefficients}) implies that surface integrals are well-defined in the ordinary sense. We note that $\Gamma$ and hence the CP potential is independent of the values of the coefficients $\mathbb{C}^{i}$, $\mathbb{C}^{e}$. However, at any finite order of the MSE, the potential does depend on the chosen coefficients, which implies that the rate of convergence of the MSE depends in general on this choice.

As a consequence of the two properties (1) and (2), the CP potential for bodies with smooth surfaces can be computed by ordinary multi-dimensional numerical integration, order by order in the number of scatterings. Hence, no knowledge of the scattering amplitude (T-matrix) of the body is required. This makes the approach much more widely applicable as the usual T-matrix based scattering approach. Compared to previous fully numerical approaches based on the boundary element method (BEM) \cite{reid2013}, here the numerical complexity is much reduced. The use of a set of surface basis functions (such as RWG basis functions \cite{Rao:1982aa}), required for the BEM \cite{reid2013}, is not needed in our approach. In fact, the basis functions are essential in the BEM to define a matrix version of the surface operator in order to be able to invert it. Also, proper basis functions are required in the BEM to deal with a stronger singularity in the surface operator, and to obtain well-defined matrix elements by numerical integration over the mesh elements.

\section{Exact Potential for a sharp PEC Wedge}
\label{sec:PEC_wedge}

For a sharp wedge made of a PEC, the Green function can be expressed in terms cylindrical wave functions of fractional order. Using the T-matrix formulation, one can obtain the exact expression for the CP potential $U_\text{PEC}(d,\phi)$ for a particle with electric polarizability $\alpha$. It is given by \cite{Deutsch:1979aa,Brevik:1998ab,Graham:2023aa}
\begin{equation}
\label{eq:CP_wedge_PEC}
U_\text{PEC}(d,\phi) = -\frac{\hbar c \, \alpha}{360 \pi\,  d^4}
\left[  \frac{135 p^4}{ \sin^4\left(p \left(\theta -\phi + \pi/2\right)\right)}
- \frac{90 \left(p^2-1\right) p^2}{ \sin^2\left(p \left(\theta -\phi +\pi/2\right)\right)} -p^4 - 10 p^2+11 \right]
\end{equation}
where $d$ and $\phi$ specify the position of the particle relative to the tip of the sharp wedge (see Fig.~\ref{fig:2} with $R=0$) and $p=\pi/(\pi+2\theta)$. We note that this result holds for $-\pi/2<\theta\le \pi/2$, i.e., both for a convex and a concave wedge. For $\theta=0$ the potential reduces to the CP potential $U_\text{PEC}(d,\phi) = - \,(3/8)\, \hbar c \alpha /d^4_\perp$ of a planar surface with perpendicular particle distance $d_\perp=d\cos\phi$. For any value of the angle $\theta$ and for any finite separation $d$, when the particle approaches the surface, i.e., $\phi \to \pi/2+\theta$, the potential reduces again to that of a planar surface. This can be seen by expanding the sinus of the first term in Eq.~(\ref{eq:CP_wedge_PEC}) to lowest order.

\section{Dielectric Wedge}

\begin{figure}[h]
\includegraphics[width=0.95\textwidth]{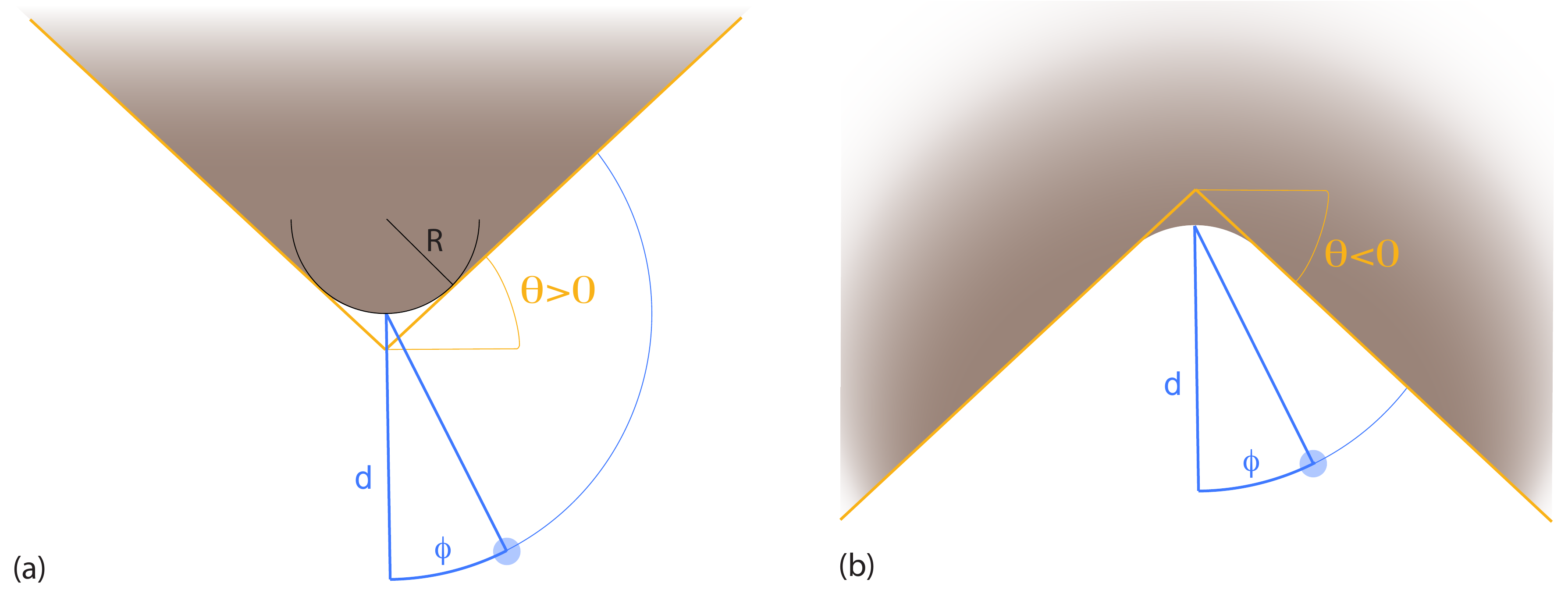} 
\caption{\label{fig:2}
Geometry of the dielectric wedge and the polarizable particle (blue) which is located at a distance $d$ from the edge of the wedge and at an angle $\phi$ with the symmetry axis of the wedge. The wedge can be convex (angle $\theta>0$ or opening angle $<\pi$) or concave (angle $\theta<0$ or opening angle $>\pi$).}
\end{figure}

As stated before, for a dielectric wedge the exact Green function is unknown, and hence the CP potential cannot be computed by the Green function method employed for the case of a PEC. Here, we use the MSE outlined in Sec.~\ref{sec:MSE}. For a surface with a sharp edge, some components of $K_{jj'}^{(pq)}({\bf u},{\bf u}')={\bf t}_j({\bf u}){\mathbb K}^{(pq)}({\bf u},{\bf u}'){\bf t}_{j'}({\bf u}')$, $p,q\in \{ E,H \}$, diverge more strongly than the inverse distance $|{\bf u}-{\bf u}'|$ when ${\bf u}$ is located on the edge, and ${\bf u}'$ approaches ${\bf u}$. Here ${\bf t}_j$ are the two tangential surface vector fields along the edge ($j=z$) and perpendicular to it ($j=\perp$). For a dielectric material with $\mu_1=1$ and $\epsilon_0=1, \, \mu_0=1$, the most divergent element is  $K^{(HH)}_{zz}\sim |{\bf u}-{\bf u}'|^{-2}$ if ${\bf u'}$ approaches the edge at a finite angle with the axis of the edge. If ${\bf u}$ and ${\bf u'}$ approach each other along the edge, the singularity is at most $\sim |{\bf u}-{\bf u}'|^{-1}$ for all elements of ${\mathbb K}$. Due to the behavior of $K^{(HH)}_{zz}$, the surface operator ${\mathbb K}$ can be integrated in the ordinary sense (without using principal values or distributions) only if the edge is smoothed.
Hence, we consider here a wedge with a smooth edge with a radius of curvature $R$ so that the surface normal vector is continuous along the entire surface. The configuration and its parametrization is shown in Fig.~\ref{fig:2}. For a concave wedge with $\theta<0$ we have to choose $R<0$. The problem involves two geometric length scales, $d$ and $R$. As the CP potential must be proportional to $\hbar c \, \alpha$ it must decay as $1/d^4$ for all distances $d$ when the dielectric permittivity $\epsilon_1$ is constant, as assumed here.  Therefore, the potential can be written as $U({\bf r}_0) = -  ( \hbar c \, \alpha/d^4 ) \Upsilon_{R/d} ({\bf r}_0)$ with a dimensionless function $\Upsilon_{R/d}$ which depends only on the ratio $R/d$. It is expected that this function converges to the potential of a sharp wedge, $\Upsilon_{0}$, when the MSE approach is applied to decreasingly smaller $R/d$. 

We introduce a few lengths which are convenient in the discussions below. The oriented separation between the tip of the sharp wedge and the tip of the smooth wedge is given by 
\begin{equation}
\delta = R \left( \frac{1}{\cos\theta} - 1\right) \, 
\end{equation}
which can be positive or negative, depending on the sign of $R$. The 
shortest normal distance of the particle from the surface of the smooth wedge is
\begin{equation}
d_\perp (\phi) = \left\{ \begin{matrix} \sqrt{d^2+R^2+2dR\cos\phi} - R & \text{if} \quad \phi < \theta + \arcsin \left(\frac{R}{d}\sin \theta\right) \\ d\cos(\theta-\phi)+R(\cos\theta -1) & \text{if} \quad \phi \ge  \theta + \arcsin \left(\frac{R}{d}\sin \theta\right)\end{matrix} \right.
\end{equation}
which holds both for positive and negative $R$, $\theta$. For the comparison with the potential of the sharp wedge ($R=0$) in Eq.~(\ref{eq:CP_wedge_PEC}), we shall also need the position of the particle relative to the {\it sharp} wedge. When the position of the particle relative to the smooth wedge is given by the surface distance $d$ and angle $\phi$ (see Fig.~\ref{fig:2}), then its polar coordinates relative to the edge of the sharp wedge are given by
\begin{align}
\label{eq:polar_coord_sharp_wedge}
d_s(\phi) & = \sqrt{d^2-2dR\cos(\phi)\left(\frac{1}{\cos\theta}-1\right)+R^2\left(\frac{1}{\cos\theta}-1\right)^2}\\
\phi_s(\phi)  & =  \arctan(R+d\cos\phi-R/\cos\theta,d \sin\phi) \,.
\end{align}
where $\arctan(x,y)$ gives the arc tangent of $y/x$  taking into account which quadrant the point $(x,y)$ is in.

Finally, the CP potential for a dielectric planar surface with $\mu_0=\mu_1=1$ is \cite{parsegian}
\begin{equation}
\label{eq:CP_planar}
U_\text{plate}(d) = - \hbar c \, \alpha \int_0^\infty \frac{d\kappa}{2\pi} \int_0^\infty dk \frac{k}{\epsilon_0 s_0}\left[ (\epsilon_0 \kappa^2+2k^2)\frac{\epsilon_1 s_0-\epsilon_0 s_1}{\epsilon_1 s_0+\epsilon_0 s_1} - \epsilon_0 \kappa^2 \frac{s_0-s_1}{s_0+s_1}\right] e^{-2s_0 d}
\end{equation}
with $s_0=\sqrt{\epsilon_0 \kappa^2+k^2}$ and $s_1=\sqrt{\epsilon_1 \kappa^2+k^2}$. Rescaling shows that $U_\text{plate}(d) = - \Upsilon_\text{plate} \, \hbar c \, \alpha/d^4$ where the coefficient $\Upsilon_\text{plate}$  depends only on $\epsilon_0$ and $\epsilon_1$ (if the latter are constant). This potential provides the proximity  approximation  for a non-planar surface like the wedge. It is given by the planar surface potential at the minimal distance $d_\perp$ of the particle from the surface, $U_\text{plate}(d_\perp)$.

\subsection{Numerical approach}

For the smoothed wedge, the kernel ${\mathbb K}$ is only weakly singular (proportional to the inverse distance between the surface points) so that it can be integrated over the surface without any regularization techniques and the use of basis functions. 
In our MSE, no operator needs to be inverted directly, so that the matrix elements of ${\mathbb K}$ relative to some basis are not required. The computation of the latter is the most difficult and computational expensive step in BEM's for scattering problems, as the computation of each matrix element involves a 4-dimensional numerical integration (2 surface integrals over the two basis functions). To carry out the surface integrations in Eq.~(\ref{eq:CP_1st_order}) we perform a change of variables to the distance vector between the two surface points of the operator $\mathbb{K}$, i.e., the new integration runs over ${\bf u}$, ${\bf \delta u}={\bf u}-{\bf u}'$, ${\bf \delta u}'={\bf u}'-{\bf u}''$ for the CP potential to second order in $\mathbb{K}$. Next, we introduce polar coordinates for these 2D integration variables. The Jacobian of this transformation cancels the inverse distance singularity of the operator $\mathbb{K}$ when ${\bf \delta u}, {\bf \delta u}' \to {\bf 0}$. We show below that a MSE to quadratic order in $\mathbb{K}$ is sufficient to obtain accurate estimates of the CP potential for all considered values of $\epsilon_1$. At this order, we need to perform 7-dimensional numerical integrations, including the integral over $\kappa$. Since $\mathbb{K}$ decays exponentially for large ${\bf \delta u}$, the integrals are in general well-behaved, and numerical integration poses no problem. We performed the numerical integrations with a requested relative error of $1\%$ or smaller but for smooth integrands the estimated error is almost always conservative. 

It turns out that the 2nd order MSE contains already a remarkable amount of information on the exact form of the CP potential. This can be seen by applying a non-linear series acceleration method, known as Shanks transformation \cite{Shanks:1955rq}. When we denote by $\delta U_l({\bf r}_0)$ the contribution of the term $\sim \mathbb{K}^l$ in the MSE of Eq.~(\ref{eq:CP_1st_order}), then the exact CP potential is given by $U({\bf r}_0)=\sum_{l=0}^\infty \delta U_l({\bf r}_0)$. It is useful to consider the partial sums $U_l({\bf r}_0)=\sum_{n=0}^l \delta U_n({\bf r}_0)$ which are the $l^\text{th}$ order MSE approximations to the CP potential. Then the Shanks transformation $S(U_l)$ of the sequence $U_l$ is the new sequence defined by
\begin{equation}
\label{eq:shanks}
S(U_l) = \frac{U_{l+1}U_{l-1}-U_l^2}{U_{l+1}-2U_l+U_{l-1}}
\end{equation}
for $l\ge 1$. This sequence $S(U_l)$ often converges much more rapidly than the sequence $U_l$. This transformation has its origin in the observation that partial sums often behave approximately as $U_l = U +\alpha q^l$ with $|q|<1$ for larger $l$. We shall observe below that $S(U_1)$, and hence the first three MSE orders $U_l$, $(l=0,1,2)$, provide very good estimates of the exact CP potential.  For a wedge with a large dielectric constant, $\epsilon=100$, however, partial sums involving  $\delta U_l({\bf r}_0)$ for even and odd $l$ have to be considered separately. The reason for that is related to the fact that for a PEC the  $\delta U_l({\bf r}_0)$ for odd $l$ vanish \cite{balian1978}. Hence, for sufficiently large $\epsilon$ the odd order contributions become orders of magnitude smaller than the even order contributions. However, the above procedure can still be applied to the sequences of partial sums of only even and only odd contributions, and the CP potential is given by the sum of the two Shanks transformed sequences.

\subsection{Results}

\begin{figure}[h]
\includegraphics[width=1.0\textwidth]{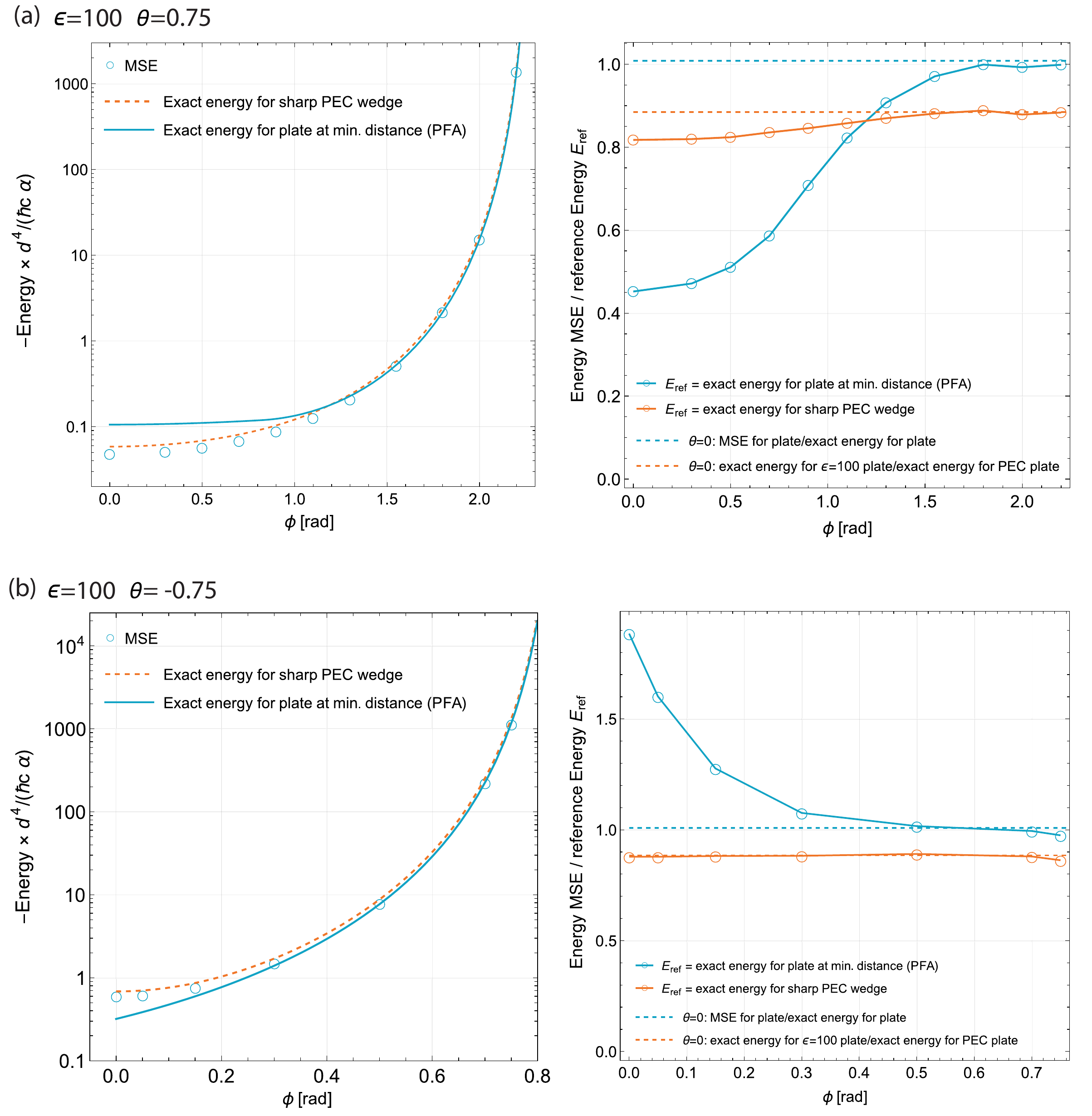} 
\caption{\label{fig:3}
Casimir-Polder potential for a wedge with $\epsilon_1=100$ for fixed smoothing radius $|R|/d=0.1$. (a) Convex case ($\theta=0.75$), (b) Concave case ($\theta=-0.75$). The panels in the left column show the (rescaled) potential as function of the angular position of the particle. The panels in the right column show the MSE result for the potential divided by a reference energy $E_\text{ref}$ which is the proximity estimate (PFA) or the potential for a sharp PEC wedge. Also shown are ratios for plates ($\theta=0$) as dashed lines: the MSE potential divided by the exact potential (for $\epsilon_1$), and the exact potential (for $\epsilon_1$) divided by the exact potential for a PEC.}
\end{figure}

\begin{figure}[h]
\includegraphics[width=1.0\textwidth]{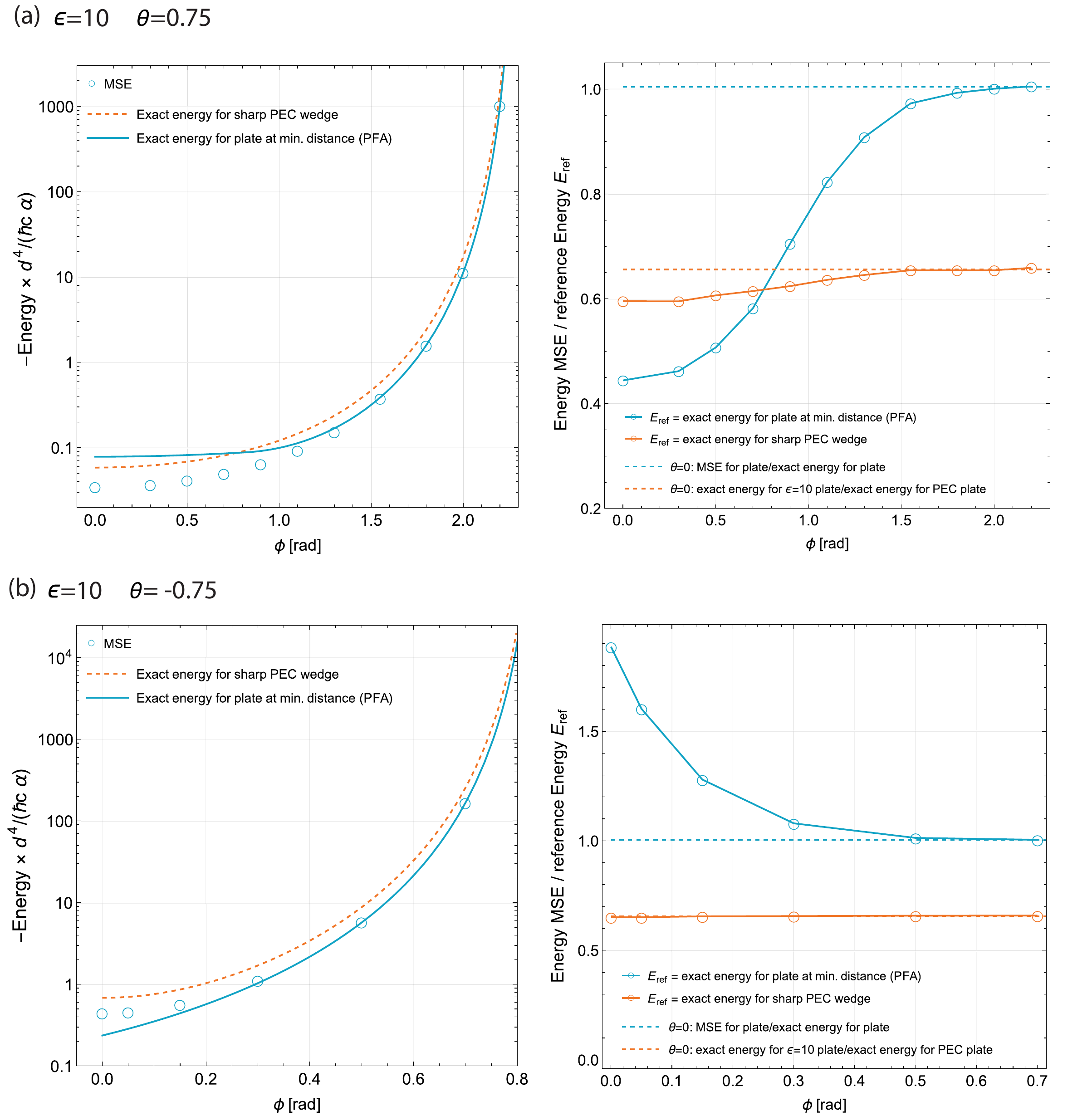} 
\caption{\label{fig:4}
Casimir-Polder potential for a wedge with $\epsilon_1=10$ for fixed smoothing radius $|R|/d=0.1$. (a) Convex case ($\theta=0.75$), (b) Concave case ($\theta=-0.75$). For the meaning of the curves, see Fig.~\ref{fig:3}.}
\end{figure}

\begin{figure}[h]
\includegraphics[width=1.0\textwidth]{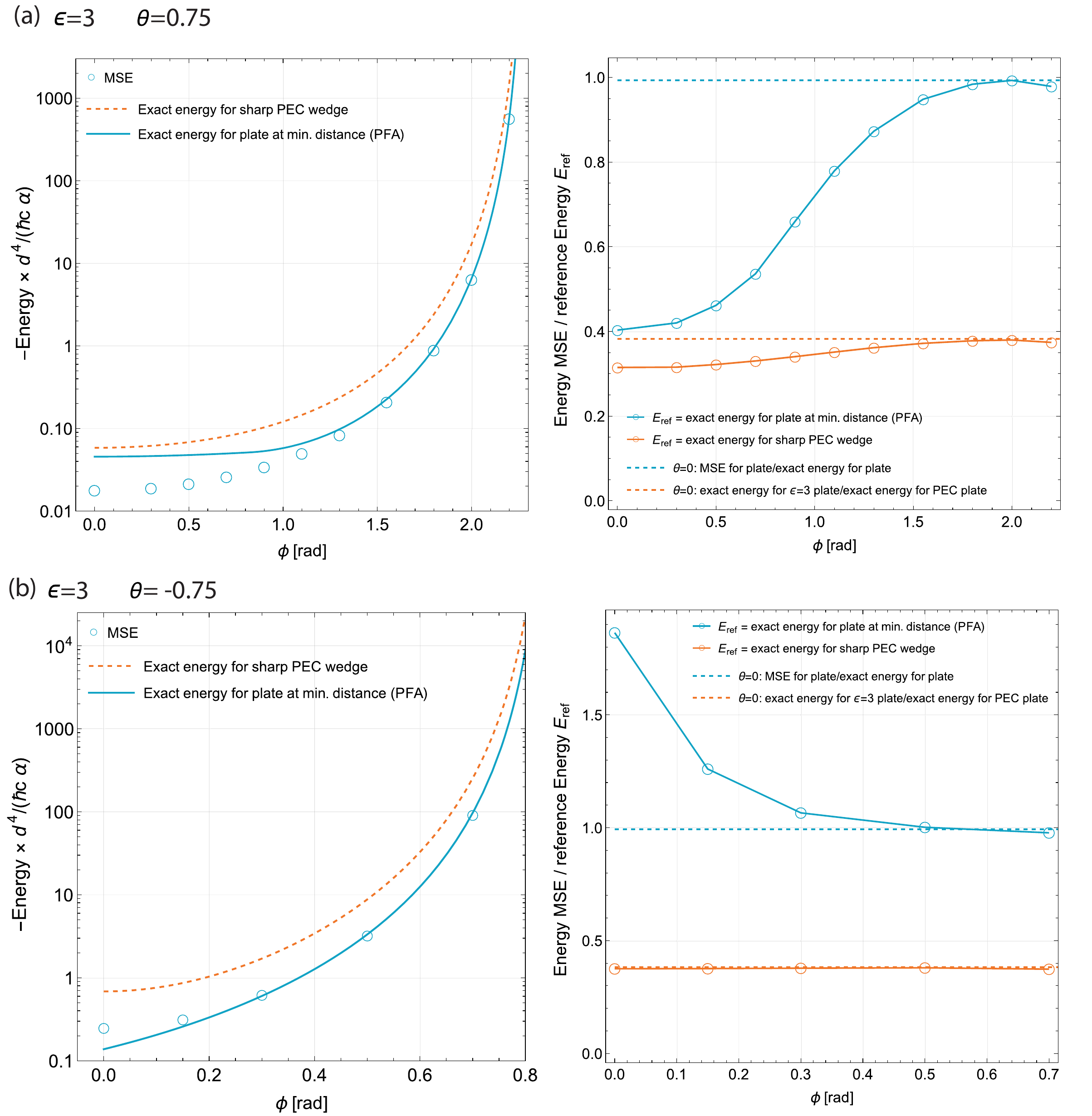} 
\caption{\label{fig:5}
Casimir-Polder potential for a wedge with $\epsilon_1=3$ for fixed smoothing radius $|R|/d=0.1$. (a) Convex case ($\theta=0.75$), (b) Concave case ($\theta=-0.75$). For the meaning of the curves, see Fig.~\ref{fig:3}.}
\end{figure}

We consider three different constant electric permittivities for the wedge, which is placed in vacuum: $\epsilon_0=1$, $\mu_0=1$, $\mu_1=1$, and $\epsilon_1=3,10$ and $100$. We compute the CP potential for a convex smooth wedge with $\theta=0.75$ and for a concave smooth wedge with $\theta=-0.75$. The radius of curvature of the edge is set to $R/d=\text{sgn}(\theta)\, 0.1$, see Fig.~\ref{fig:2}. The numerical integrations were performed to second MSE order, i.e, to order  $\mathbb{K}^2$, and then were Shanks transformed with the final estimate for the CP potential given by $S(U_1)$, see Eq.~(\ref{eq:shanks}). Before we consider the wedge set-up, it is instructive to apply our MSE procedure to the CP potential of a planar surface. The exact potential is known, and given by Eq.~(\ref{eq:CP_planar}). The amplitude of the potential, $\Upsilon_\text{plate}$, is a numerical constant which depends only on the dielectric constant (see text below Eq.~\ref{eq:CP_planar})). Our MSE estimate for $\epsilon_1=100$ is $\Upsilon_\text{plate}^\text{MSE}=0.1065$ (exact value $\Upsilon_\text{plate}=0.1056$, $0.85\%$ error), for $\epsilon_1=10$ it is $\Upsilon_\text{plate}^\text{MSE}=0.0786$ (exact value $\Upsilon_\text{plate}=0.0783$, $0.44\%$ error), and for $\epsilon_1=3$ it is $\Upsilon_\text{plate}^\text{MSE}=0.0453$ (exact value $\Upsilon_\text{plate}=0.0456$, $0.73\%$ error). For a second order MSE with convergence acceleration this agreement over a wide range of dielectric constants is rather astonishing. 

Having established the reliability of the MSE,  we now turn to the wedge geometry. Our MSE estimates for the rescaled CP potential $\Upsilon_{R/d}(\phi)$ are shown in Figs.~\ref{fig:3}--\ref{fig:5}, both for the convex wedge [panels (a)] and for the concave wedge [panels (b)]. Note that the rescaled  potential only depends on the angular position $\phi$ of the particle as $R/d$ has been fixed. For comparison, the left columns of these figures show also the exact CP potential for a sharp PEC wedge, $U_\text{PEC}(d_s(\phi),\phi_s(\phi))$, and the exact CP potential for a dielectric plate at shortest wedge-particle distance, $U_\text{plate}(d_\perp(\phi))$, which can be regarded as a proximity approximation, known as PFA for the Casimir force. 
Here it is important to take the potential of the sharp PEC wedge at the  particle coordinates $(d_s(\phi),\phi_s(\phi))$ relative to its edge in order to compare equivalent geometries. The CP potential always attracts the particle towards the surface, as its negative amplitude increases when $\phi$ increases. 
Both for the convex and for the concave wedge, the CP potential shows an angular dependence on $\phi$ which is similar to that of a sharp PEC wedge but with a reduced amplitude which decreases with decreasing value of $\epsilon_1$. 
For the convex wedge, for angles $\phi$ larger than about $1.5$rad (which is at about $65\%$ of the possible range $0\le \phi <\pi/2+\theta$), the potential is well approximated by the potential of a plate placed at the shortest wedge-particle distance. For the concave wedge, the same holds true for angles larger than about $0.5$rad (which is at about $61\%$ of the possible range). Hence for those larger angles $\phi$ the particle mainly senses a single planar surface and the proximity approximation is justified.

To make the relation of the MSE estimate to known results and approximations more visible, we consider now certain quotients of potentials.
The right columns of Figs.~\ref{fig:3}--\ref{fig:5} display the quotient of our MSE estimates of the CP potential and the exact CP potential for a sharp PEC wedge, $U_\text{PEC}(d_s(\phi),\phi_s(\phi))$, and the exact CP potential for a dielectric plate at shortest wedge-particle distance, $U_\text{plate}(d_\perp(\phi))$, respectively. Also shown, as dashed lines, 
are the quotient of our MSE estimate for a plate and the corresponding exact result, and the quotient of the exact potentials for a plate with the considered value of $\epsilon_1$ and for a PEC plate. 
These figures demonstrate the following:
(1) We observe in these plots even more clearly that the proximity approximation becomes accurate when the particle approaches the planar wall of the wedge for increasing angle $\phi$. (2) The MSE estimate of the CP potential in this limit agrees very well with the MSE estimate for a planar surface, which we found to be in excellent agreement with the known exact result, see discussion above. (3) For a convex wedge the proximity approximation overestimates the actual CP potential while for a concave wedge this approximation underestimates the actual CP potential. For all three considered dielectric constants, in the convex case the actual potential is between $40\%$ and $45\%$ of the proximity approximation for $\phi=0$. In the concave case the actual potential is about $1.9$ times the proximity approximation for $\phi=0$. The latter result suggests that the potential is approximately the sum of the two equal contributions from both planar surfaces which are at the same distance from the particle for $\phi=0$. (4) The most interesting observation follows from a comparison with the exact potential for a {\it sharp} PEC wedge, $U_\text{PEC}(d_s(\phi),\phi_s(\phi))$. For a concave smooth dielectric wedge we find that the CP potential is given by the potential for a {\it sharp} PEC wedge times an overall, geometry-independent factor which is the exactly known quotient $\Upsilon_\text{plate}(\epsilon_1)/\Upsilon_\text{plate}^\text{PEC}$ of the potential for a dielectric plate and a PEC plate with $\Upsilon_\text{plate}^\text{PEC}=3/8$. This factor describes the reduction of the potential for a dielectric compared to a PEC. Interestingly, we find that this result holds for all considered values of $\epsilon_1$, within the numerical precision of our results, see the solid and dashed orange curves in the right columns of Figs.~\ref{fig:3}--\ref{fig:5}. This implies that the finite curvature $R/d=-0.1$ of the edge does not change the $\phi$-dependence of the potential compared to that of a sharp concave wedge, within the numerical precision of our results. 
 For a convex wedge, however, we observe a different geometry dependence. Only for larger $\phi$, when the particle approaches the planar surface of the wedge, the CP potential of the smooth dielectric wedge is given again by $U_\text{red}(\phi)=(\Upsilon_\text{plate}(\epsilon_1)/\Upsilon_\text{plate}^\text{PEC} )\, U_\text{PEC}(d_s(\phi),\phi_s(\phi))$. 
For smaller $\phi$, when the particle is located more closely to the edge of the wedge, there is an additional, $\phi$-dependent reduction of the CP potential compared to $U_\text{red}(\phi)$. For $\phi=0$ this reduction is $8\%$ for $\epsilon_1=100$, $9\%$ for $\epsilon_1=10$ and $18\%$ for $\epsilon_1=3$. This is visible in the right columns of Figs.~\ref{fig:3}--\ref{fig:5} by a deviation of the orange curves from the dashed orange lines. 

It is desirable to have a physical explanation for this remarkable relation between the potential for a smooth dielectric wedge and that of a sharp PEC wedge. For a concave wedge, most of the interaction arises from the two planar sides of the wedge. Therefore, the overall geometry dependence is given by that of sharp PEC wedge, reduced by the factor which described the reduction of the CP interaction for a planar wall relative to a PEC wall.
The effect of the edge, and hence its curvature, is negligible even though some dielectric material is added due to the finite curvature, see Fig.~\ref{fig:2}(b). Moreover, for an edge with an opening angle larger than $\pi$ the field fluctuations are reduced compared to a planar surface as less surface charges are induced in its vicinity. For a convex wedge the situation is different.
Only when the particle is located close to one of the side walls of the wedge, i.e., for sufficiently large $\phi$, the above argument for a concave wedge applies. However, if the particle is located close to the edge, it probes field fluctuations that are different from a planar surface. Hence, the main reason for the observed additional reduction of the CP interaction relative to a sharp PEC wedge should be the removed material due the finite curvature of the edge, see Fig.~\ref{fig:2}(a). In addition, for an edge with an opening angle smaller than $\pi$ the field fluctuations are increased compared to a planar surface as more surface charges are induced in its vicinity, increasing the relative importance of the edge compared to the planar parts of the wedge.

\section{Conclusion}

In this work, for the first time, we have implemented a basis-free multiple scattering expansion for Casimir-Polder interactions between a polarizable particle and a dielectric surface for which the scattering amplitude (T-matrix) is not known. Combining a low order multiple scattering expansion with a non-linear convergence acceleration method we obtained 
high precision estimates of the Casimir-Polder potential, 
for a wide range of dielectric constants. This is a substantial progress compared to T-matrix based scattering methods and
fully numerical boundary element methods with the latter requiring a surface discretization and the use of suitable basis functions. The numerical complexity is much reduced for our approach. We observed the remarkable result that Casimir-Polder potential for a smooth concave dielectric wedge is determined by the potential for a sharp PEC wedge times the same reduction factor which applies to a simple planar surface. Our approach is not specific to the quasi-2D geometry of a smooth wedge but can be applied to general, smooth 3D surface shapes since it requires only the numerical integration of an explicitly known function (the trace of scattering operators) over the surface.

\begin{acknowledgments}
Discussions with G.~Bimonte are acknowledged.
\end{acknowledgments}

 \clearpage 



\end{document}